\documentstyle[prd,aps]{revtex}

\newcommand{\bea}{\begin{eqnarray}}
\newcommand{\eea}{\end{eqnarray}}

\begin{document}
%%%%%%%%%%%%%%%%%%%%%%%%%%%%%%%%%%%%%%%%%%%%%%%%%%%%%%%%%%%%%%%
\draft
%  For 2 column format.
\twocolumn[\hsize\textwidth\columnwidth\hsize\csname
@twocolumnfalse\endcsname

%%%%%%%%%%%%%%%%%%%%%%%%%%%%%%%%%%%%%%%%%%%%%%%%%%%%%%%%%%%%%%%
\title{Second-order perturbations of a zero-pressure cosmological
       medium: \\
       comoving vs. synchronous gauge}

\author{Jai-chan Hwang${}^{(a)}$ and Hyerim Noh${}^{(b)}$}
\address{${}^{(a)}$ Department of Astronomy and Atmospheric Sciences,
                    Kyungpook National University, Taegu, Korea \\
         ${}^{(b)}$ Korea Astronomy and Space Science Institute,
                    Daejon, Korea \\
         E-mails: ${}^{(a)}$jchan@knu.ac.kr, ${}^{(b)}$hr@kasi.re.kr
         }
%\date{\today}
\maketitle

%%%%%%%%%%%%%%%%%%%%%%%%%%%%%%%%%%%%%%%%%%%%%%%%%%%%%%%%%%%%%%%
\begin{abstract}

Except for the presence of gravitational wave source term, the
relativistic perturbation equations of a zero-pressure irrotational
fluid in a flat Friedmann world model coincide exactly with the
Newtonian ones to the second order in perturbations. Such a
relativistic-Newtonian correspondence is available in a special
gauge condition (the comoving gauge) in which all the variables are
equivalently gauge invariant. In this work we compare our results
with the ones in the synchronous gauge which has been used often in
the literature. Although the final equations look simpler in the
synchronous gauge, the variables have remnant gauge modes. Except
for the presence of the gauge mode for the perturbed order
variables, however, the equations in the synchronous gauge are gauge
invariant and can be exactly identified as the Newtonian
hydrodynamic equations in the Lagrangian frame. In this regard, the
relativistic equations to the second order in the comoving gauge are
the same as the Newtonian hydrodynamic equations in the Eulerian
frame. We resolve several issues related to the two gauge conditions
often to fully nonlinear orders in perturbations.

\end{abstract}
%%%%%%%%%%%%%%%%%%%%%%%%%%%%%%%%%%%%%%%%%%%%%%%%%%%%%%%%%%%%%%%
%\noindent \pacs{PACS numbers: 04.50.+h, 04.62.+v, 98.80.-k,
%98.80.Hw}

%%%%%%%%%%%%%%%%%%%%%%%%%%%%%%%%%%%%%%%%%%%%%%%%%%%%%%%%%%%%%%%
%  For 2 column format.
\vskip2pc]

%%%%%%%%%%%%%%%%%%%%%%%%%%%%%%%%%%%%%%%%%%%%%%%%%%%%%%%%%%%%%%%
%
% Introduction
%
%%%%%%%%%%%%%%%%%%%%%%%%%%%%%%%%%%%%%%%%%%%%%%%%%%%%%%%%%%%%%%%

%%%%%%%%%%%%%%%%%%%%%%%%%%%%%%%%%%%%%%%%%%%%%%%%%%%%%%%%%%%%%%%
\section{Introduction}
                                          \label{sec:Introduction}

The general relativistic cosmological linear perturbation theory was
first developed by Lifshitz in 1946 \cite{Lifshitz-1946}. Lifshitz
took the synchronous gauge condition in which the perturbations of
the time-time part and the space-time part of the metric tensor are
equal to zeros; this gauge condition can be taken to fully nonlinear
order without losing any physical degree of freedom \cite{LL}. The
synchronous gauge condition has been popular in the cosmological
perturbation literature despite the complicating fact that, except
for the zero-pressure case, there are remnant gauge modes for both
the spatial and temporal gauge conditions. There exist other spatial
and temporal gauge conditions which fix the gauge transformation
property completely in general situation, thus without any remaining
gauge mode \cite{Harrison-1967,Field-Shepley-1968,Nariai-1969}. This
point was clarified by Bardeen \cite{Bardeen-1980,Bardeen-1988}. In
a zero-pressure medium the density perturbation equation in the
synchronous gauge coincides with the one in the comoving gauge
\cite{Lifshitz-1946,Nariai-1969}. The density perturbation equation
in the comoving gauge condition is known to resemble the Newtonian
equation most closely \cite{Nariai-1969,Bardeen-1980}, and the
equations coincide in the zero-pressure case
\cite{Lifshitz-1946,Bonnor-1957}. Thus, in the zero-pressure case
the density perturbation equation in the synchronous gauge coincides
with the Newtonian one to the linear order
\cite{Lifshitz-1946,Bonnor-1957}.

The synchronous gauge was also used in the nonlinear perturbation
studies \cite{Tomita}, and Kasai \cite{Kasai-1992} has derived
second-order differential equations for density perturbation which
is valid to fully nonlinear order. Although, such an equation in the
synchronous gauge naturally has proper linear limit which
corresponds to the Newtonian equation, it has been unclear whether
such a correspondence continues to the nonlinear situation.
Recently, we have successfully shown an exact relativistic-Newtonian
correspondence of scalar-type perturbations to the second order
based on the comoving gauge
\cite{NL,second-order-CQG,second-order-PRD}. In the zero-pressure
case our comoving gauge condition differs from the conventional
synchronous gauge in the spatial gauge condition. In this work we
will investigate the case in the original synchronous gauge. We will
show that although the equations in the synchronous gauge look
simpler than the ones in the comoving gauge, the variables still
have remaining (spurious) spatial gauge mode to the second order.
The equations in the synchronous gauge, however, are gauge invariant
and can be identified as the Newtonian hydrodynamic equations in the
Lagrangian frame. Whereas, the equations in the comoving gauge can
be identified as the Newtonian hydrodynamic equations in the
Eulerian frame.

Results in Secs. \ref{sec:NL} and the Appendices are valid to fully
nonlinear order in perturbations, and unless mentioned otherwise
results in the remaining sections are valid to the second order in
perturbations. We closely follow notations used in
\cite{NL,second-order-CQG,second-order-PRD}. We set $c \equiv 1$.

%%%%%%%%%%%%%%%%%%%%%%%%%%%%%%%%%%%%%%%%%%%%%%%%%%%%%%%%%%%%%%%
\section{Fully nonlinear perturbations}
                                                 \label{sec:NL}

The energy conservation equation and the Raychaudhury equation give
\cite{covariant,second-order-CQG,second-order-PRD} \bea
   & & \tilde {\dot {\tilde \mu}} + \tilde \mu \tilde \theta = 0,
   \label{covariant-eq1} \\
   & & \tilde {\dot {\tilde \theta}} + \frac{1}{3} \tilde \theta^2
       + \tilde \sigma^{ab} \tilde \sigma_{ab}
       - \tilde \omega^{ab} \tilde \omega_{ab}
       + 4 \pi G \tilde \mu - \Lambda = 0,
   \label{covariant-eq2}
\eea where $\tilde \theta \equiv \tilde u^a_{\;\; ;a}$ is an
expansion scalar based on a fluid four-vector $\tilde u_a$; $\tilde
\sigma_{ab}$ and $\tilde \omega_{ab}$ are the shear and the rotation
tensors based on $\tilde u_a$, respectively; tildes indicate the
covariant quantities and the Latin indices indicate spacetime
components. An overdot with tilde is a covariant derivative along
the $\tilde u_a$ vector, e.g., $\tilde {\dot {\tilde \mu}} \equiv
\tilde \mu_{,a} \tilde u^a$. By combining these equations we have
\bea
   & & \left( \frac{\tilde {\dot {\tilde \mu}}}{\tilde \mu}
       \right)^{\tilde \cdot}
       - \frac{1}{3}
       \left( \frac{\tilde {\dot {\tilde \mu}}}{\tilde \mu} \right)^2
       - \tilde \sigma^{ab} \tilde \sigma_{ab}
       + \tilde \omega^{ab} \tilde \omega_{ab}
       - 4 \pi G \tilde \mu
       + \Lambda
       = 0.
   \nonumber \\
   \label{covariant-eq3}
\eea Equations (\ref{covariant-eq1})-(\ref{covariant-eq3}) are fully
nonlinear and covariant, thus valid to all orders in perturbations.

%%%%%%%%%%%%%%%%%%%%%%%%%%%%%%%%%%%%%%%%%%%%%%%%%%%%%%%%%%%%%%%
\subsection{Temporal Comoving Gauge}

In this work we will {\it assume} an irrotational fluid, thus
$\tilde \omega_{ab} \equiv 0$. We will consider two different gauge
conditions. In both gauge conditions we will have \bea
   & & \tilde u_\alpha = 0,
\eea due to a common temporal gauge condition together with the
irrotational condition; the Greek indices indicate space components.
If we introduce the spatial part of the four-vector as \bea
   & & \tilde u_\alpha \equiv a \left( - \hat v_{,\alpha}
       + \hat v^{(v)}_\alpha \right),
   \label{u-def}
\eea where $\hat v^{(v)}_\alpha$ is a vector-type perturbation (thus
transverse), the irrotational condition sets $\hat v^{(v)}_\alpha
\equiv 0$ and our temporal comoving gauge sets $\hat v \equiv 0$.
Since $\tilde u_\alpha = 0$ the fluid four-vector in this gauge
coincides with the {\it normal} frame four-vector $\tilde n_a$ with
$\tilde n_\alpha \equiv 0$. Notice that our temporal comoving gauge
condition $\hat v \equiv 0$ (together with the irrotational
condition) implies $\tilde u_\alpha = 0$. This {\it differs} from
the ordinarily known {\it comoving} frame condition which sets
$\tilde u^\alpha \equiv 0$ \cite{Taub-1978}. In our case the
normalized ($\tilde u^a \tilde u_a \equiv -1$) fluid four-vector
$\tilde u_a$ becomes \bea
   & & \tilde u_0 = - {1 \over \sqrt{- \tilde g^{00}}}, \quad
       \tilde u_\alpha \equiv 0;
   \nonumber \\
   & &
       \tilde u^0 = \sqrt{- \tilde g^{00}}, \quad
       \tilde u^\alpha = - {\tilde g^{0\alpha}
       \over \sqrt{- \tilde g^{00}}}.
   \label{u}
\eea

In the zero-pressure case the momentum conservation equation implies
$\tilde g^{00} = - 1/a^2$ where $a$ is the cosmic scale factor of
the Friedmann background world model. In the ADM approach
\cite{ADM}, our temporal comoving gauge $\hat v = 0$ together with
the irrotational condition implies vanishing momentum vector
$J_\alpha \equiv - \tilde n_b \tilde T^b_\alpha = 0$. The ADM
momentum conservation equation in Eq.\ (13) of \cite{NL} gives
$N_{,\alpha} = 0$ where $\tilde g^{00} \equiv -1/N^2$, thus $N =
N(t)$. In another way, since the acceleration vector $\tilde
a_\alpha \equiv \tilde u_{\alpha ;b} \tilde u^b =
(\ln{N})_{,\alpha}$ vanishes (i.e., geodesic flow) for the
zero-pressure irrotational flow, we have $N = N(t)$; see Eqs.\ (27)
and (42) of \cite{NL}. Without losing generality we can set $N =
a(t)$. Thus we have \bea
   & & \tilde g^{00} = - {1 \over a^2}.
   \label{g^00}
\eea Thus, Eq.\ (\ref{u}) becomes \bea
   & & \tilde u_0 = - a, \quad
       \tilde u_\alpha \equiv 0; \quad
       \tilde u^0 = {1 \over a}, \quad
       \tilde u^\alpha = - a \tilde g^{0\alpha},
   \label{u-pressureless}
\eea which is valid to fully nonlinear order. We can show that to
all orders in perturbations the fluid quantities are independent of
the spatial gauge condition which could affect $\tilde g^{0\alpha}$;
see the Appendix A.

%%%%%%%%%%%%%%%%%%%%%%%%%%%%%%%%%%%%%%%%%%%%%%%%%%%%%%%%%%%%%%%
\subsection{Nonlinear perturbed equations}

We introduce perturbations \bea
   & & \tilde \mu \equiv \mu + \delta \mu, \quad
       \tilde \theta \equiv 3 H - \kappa,
\eea where $H \equiv \dot a/a$ and $\delta \equiv \delta \mu/\mu$;
an overdot denotes a time derivative based on background proper-time
$t$. The $\tilde \theta$ is an expansion scalar of the fluid
four-vector which is the same as the normal four-vector because
$\tilde u_\alpha = 0$ in our case. Using Eq.\ (\ref{u-pressureless})
we have \bea
   & & \tilde {\dot {\tilde \mu}}
       = \dot \mu \left( 1 + \delta \right)
       + \mu \left( \dot \delta - {1 \over a} N^\alpha \delta_{,\alpha}
       \right),
   \nonumber \\
   & &
       \tilde {\dot {\tilde \theta}} = 3 \dot H - \left( \dot \kappa
       - {1 \over a} N^\alpha \kappa_{,\alpha} \right),
   \label{dot-mu}
\eea where $N^\alpha$ is the shift vector in the ADM notation with
$N^\alpha \equiv a^2 \tilde g^{0\alpha}$; the spatial indices of the
ADM variables are based on $h_{\alpha\beta} \equiv \tilde
g_{\alpha\beta}$.

Equations (\ref{covariant-eq1}), (\ref{covariant-eq2}) give \bea
   & & \left( \dot \mu + 3 H \mu \right) \left( 1 + \delta \right)
   \nonumber \\
   & & \quad
       + \mu \left[ \dot \delta
       - {1 \over a} \delta_{,\alpha} N^\alpha
       - \left( 1 + \delta \right) \kappa \right]
       = 0,
   \\
   & & 3 \left( \dot H + H^2 \right) + 4 \pi G \mu - \Lambda
   \nonumber \\
   & & \quad
       - \left[ \dot \kappa
       - {1 \over a} \kappa_{,\alpha} N^\alpha
       + 2 H \kappa
       - 4 \pi G \mu \delta
       - {1 \over 3} \kappa^2
       - \tilde \sigma^{ab} \tilde \sigma_{ab} \right]
   \nonumber \\
   & & \quad
       = 0.
\eea The background parts give \bea
   & & \dot \mu + 3 H \mu = 0, \quad
       3 \left( \dot H + H^2 \right) + 4 \pi G \mu - \Lambda = 0.
   \label{BG-eqs}
\eea The perturbed parts give \bea
   & & \hat {\dot \delta} = \left( 1 + \delta \right)\kappa,
   \label{NL-eq1} \\
   & & \hat {\dot \kappa} + 2 H \kappa = {1 \over 3} \kappa^2
       + \tilde \sigma^{ab} \tilde \sigma_{ab}
       + 4 \pi G \mu \delta,
   \label{NL-eq2}
\eea where $\hat {\dot \delta} \equiv \dot \delta - a^{-1}
\delta_{,\alpha} N^\alpha$. By combining these equations we have
\bea
   & & \hat {\ddot \delta}
       + 2 H \hat {\dot \delta} - 4 \pi G \mu \delta
       = 4 \pi G \mu \delta^2
       + {4 \over 3} {(\hat {\dot \delta})^2 \over 1 + \delta}
       + \left( 1 + \delta \right) \tilde \sigma^{ab} \tilde
       \sigma_{ab}.
   \nonumber \\
   \label{NL-eq3}
\eea These equations are valid to the fully nonlinear orders in
perturbations, subject only to the temporal comoving gauge
condition, the zero-pressure condition, and the irrotational
condition.

%%%%%%%%%%%%%%%%%%%%%%%%%%%%%%%%%%%%%%%%%%%%%%%%%%%%%%%%%%%%%%%
\subsection{The synchronous gauge}

Under the synchronous gauge we set $\tilde g_{0\alpha} \equiv 0$
(thus $N^\alpha \equiv 0$) using the spatial gauge condition
(together with the irrotational condition), thus \bea
   & & \tilde {\dot {\tilde \mu}}
       = \hat {\dot {\tilde \mu}}
       = {\dot {\tilde \mu}}.
\eea Thus, Eqs.\ (\ref{NL-eq1})-(\ref{NL-eq3}) simply give \bea
   & & \dot \delta = \left( 1 + \delta \right)\kappa,
   \label{NL-SG-eq1} \\
   & & \dot \kappa + 2 H \kappa = {1 \over 3} \kappa^2
       + \tilde \sigma^{ab} \tilde \sigma_{ab}
       + 4 \pi G \mu \delta,
   \label{NL-SG-eq2} \\
   & & \ddot \delta + 2 H \dot \delta - 4 \pi G \mu \delta
       = 4 \pi G \mu \delta^2
       + {4 \over 3} {\dot \delta^2 \over 1 + \delta}
       + \left( 1 + \delta \right) \tilde \sigma^{ab} \tilde
       \sigma_{ab},
   \nonumber \\
   \label{NL-SG-eq3}
\eea which are valid to the fully nonlinear order. Using $\Delta
\equiv \delta/(1 + \delta)$ Kasai \cite{Kasai-1992} has derived \bea
   & & \ddot \Delta + 2 H \dot \Delta - 4 \pi G \mu \Delta
       = - {2 \over 3} {\dot \Delta^2 \over 1 - \Delta}
       + \left( 1 - \Delta \right) \tilde \sigma^{ab} \tilde
       \sigma_{ab}.
   \nonumber \\
\eea To nonlinear order in perturbations the above equations are
incomplete yet because of $\tilde \sigma^{ab} \tilde \sigma_{ab}$
term. Later we will show that these equations in the synchronous
gauge differs from the equations in the comoving gauge to the second
order. Furthermore, although these equations look simple, we will
show that $\delta$ (thus $\Delta$ as well) and $\kappa$ still have
remnant gauge modes to the second order. In Sec.\ \ref{sec:SG} we
will show that to the second order the equations are gauge invariant
and can be identified with the Newtonian hydrodynamic equations in
the Lagrangian frame. In this regard, the equations in the comoving
gauge correspond to the Newtonian hydrodynamic equations in the
Eulerian frame.

%%%%%%%%%%%%%%%%%%%%%%%%%%%%%%%%%%%%%%%%%%%%%%%%%%%%%%%%%%%%%%%
\section{Second-order perturbations}

As the metric we take \bea
   & & ds^2
       = - a^2 \left( 1 + 2 \alpha \right) d \eta^2
       - 2 a^2 \beta_{,\alpha} d \eta d x^\alpha
   \nonumber \\
   & & \quad
       + a^2 \left[ g^{(3)}_{\alpha\beta} \left( 1 + 2 \varphi \right)
       + 2 \gamma_{,\alpha|\beta}
       + 2 C^{(t)}_{\alpha\beta} \right] d x^\alpha d x^\beta,
   \label{metric}
\eea where $\alpha$, $\beta$, $\gamma$ and $\varphi$ are spacetime
dependent perturbed-order variables, and $C^{(t)}_{\alpha\beta}$ is
a transverse and tracefree perturbed-order variable. Spatial indices
of perturbed order variables are based on $g^{(3)}_{\alpha\beta}$,
and a vertical bar indicates the covariant derivative based on
$g^{(3)}_{\alpha\beta}$;  $g^{(3)}_{\alpha\beta}$ could become
$\delta_{\alpha\beta}$ in a flat Friedmann background. We {\it
ignored} the transverse vector-type perturbation variables. We
introduce $\chi \equiv a ( \beta + a \dot \gamma)$. The perturbed
variables can be regarded as nonlinearly perturbed ones to any order
in perturbations.

To the second order, from Eqs.\ (55), (57) of \cite{NL} we have \bea
   & & N^\alpha = - \beta^{,\alpha},
   \nonumber \\
   & & \tilde \sigma^{ab} \tilde \sigma_{ab}
       = \bar K^\alpha_\beta \bar K^\beta_\alpha
       = {1 \over a^4} \left[ \chi^{,\alpha|\beta}
       \chi_{,\alpha|\beta}
       - {1 \over 3} \left( \Delta \chi \right)^2 \right]
   \nonumber \\
   & & \quad
       + \dot C^{(t)\alpha\beta} \left( {2 \over a^2}
       \chi_{,\alpha|\beta} + \dot C^{(t)}_{\alpha\beta} \right),
   \label{sigma}
\eea where $N^\alpha$ is evaluated to the linear order; $\bar
K^\alpha_\beta$ is a tracefree part of the extrinsic curvature. We
note that $\tilde \sigma^{ab} \tilde \sigma_{ab}$ is spatially gauge
invariant to the second order, see Sec.\ \ref{sec:gauge-issue}.

Before comparing equations in the two different spatial gauges, we
{\it compare} our $\hat v$ in Eq.\ (\ref{u-def}) with the notation
used in \cite{NL} to the second order in perturbations. In \cite{NL}
we introduced the fluid quantities based on the normal-frame vector
$\tilde n_a$ and provided the relation of fluid quantities between
the energy-frame ($E$) and the normal-frame ($N$). Our fluid
four-vector $\tilde u_a$ is based on the energy frame which sets
$\tilde q_a \equiv 0$. The energy-frame fluid four-vector is
introduced in Eq.\ (53) of \cite{NL}, and using the relations given
in Eqs.\ (87), (88) of \cite{NL} we have \bea
   & & \tilde u_\alpha
       = a \left( V_\alpha^E - B_\alpha + A B_\alpha
       + 2 V^\beta_E C_{\alpha\beta} \right)
   \nonumber \\
   & & \quad
       = a \left\{ {Q_\alpha^N \over \mu + p}
       - {1 \over (\mu + p)^2} \left[
       \left( \delta \mu + \delta p \right) Q_\alpha^N
       + Q^\beta_N \Pi_{\alpha\beta} \right] \right\}.
   \nonumber \\
\eea Using the decomposition of the normal-frame flux vector
$Q_\alpha^N \equiv (\mu + p) ( - v_{,\alpha} + v_\alpha^{(v)})$ in
Eq.\ (175) of \cite{NL} and setting $v_\alpha^{(v)} \equiv 0$ we
have \bea
   & & \hat v_{,\alpha}
       = v_{,\alpha}
       - {1 \over \mu + p} \left[
       \left( \delta \mu + \delta p \right) v_{,\alpha}
       + v^{,\beta} \Pi_{\alpha\beta} \right].
\eea Thus, the temporal comoving gauge $v \equiv 0$ in \cite{NL}
implies $\hat v = 0$ and vice versa.

%%%%%%%%%%%%%%%%%%%%%%%%%%%%%%%%%%%%%%%%%%%%%%%%%%%%%%%%%%%%%%%
\subsection{The comoving gauge}
                                                   \label{sec:CG}

In \cite{NL,second-order-CQG,second-order-PRD} we took the temporal
comoving gauge and the spatial $\gamma = 0$ gauge \bea
   & & v \equiv 0, \quad
       \gamma \equiv 0.
   \label{CG}
\eea In this work, we call this the {\it comoving} gauge. Thus, we
have $\beta = \chi/a$.

The momentum conservation equation in Eq.\ (105) of \cite{NL} gives
\bea
   & & \alpha = - {1 \over 2 a^2} \chi^{,\beta} \chi_{,\beta}.
   \label{alpha-CG}
\eea Thus, apparently, $\alpha$ does not vanish to the second order.
Later we will show that if we take $\beta = 0$ as the spatial gauge
condition instead of $\gamma = 0$, we have vanishing $\alpha$.
However, we prefer $\gamma \equiv 0$ as the spatial gauge condition
because it fixes the spatial gauge degree of freedom completely (as
long as we simultaneously take the temporal gauge which removes the
temporal gauge degree of freedom completely, like our $v = 0$), see
Sec.\ VI of \cite{NL}. Whereas, $\beta \equiv 0$ fails to fix the
spatial gauge degree of freedom completely, thus having remaining
gauge degree of freedom even after imposing the gauge condition, see
Sec.\ \ref{sec:two-gauges}.

In our gauge the fluid four-vector in Eq.\ (\ref{u-pressureless})
becomes \bea
   & & \tilde u_0 = - a , \quad
       \tilde u_\alpha = 0;
   \nonumber \\
   & &
       \tilde u^0 = {1 \over a}, \quad
       \tilde u^\alpha = \frac{1}{a^2} \chi^{,\beta}
       \left[ \left( 1 - 2 \varphi \right) \delta^\alpha_\beta
       - 2 C^{(t)\alpha}_{\;\;\;\;\;\beta} \right].
   \label{u-CG}
\eea Thus, although we prefer to call this the temporal comoving
gauge (see \cite{Bardeen-1980,Bardeen-1988}), because $\tilde
u_\alpha = 0$ and $\tilde u^\alpha \neq 0$, our fluid four-vector
corresponds to the normal four-vector rather than the comoving one.

Using Eq.\ (\ref{sigma}), Eqs.\ (\ref{NL-eq1}), (\ref{NL-eq2}) give
\bea
   & & \dot \delta
       + {1 \over a^2} \delta_{,\alpha} \chi^{,\alpha}
       - \kappa
       = \delta \kappa,
   \label{delta-eq-CG} \\
   & & \dot \kappa
       + {1 \over a^2} \kappa_{,\alpha} \chi^{,\alpha}
       + 2 H \kappa
       - 4 \pi G \mu \delta
   \nonumber \\
   & & \quad
       = \left( {1 \over a^2} \chi^{,\alpha|\beta}
       + \dot C^{(t)\alpha\beta} \right)
       \left( {1 \over a^2} \chi_{,\alpha|\beta}
       + \dot C^{(t)}_{\alpha\beta} \right).
   \label{kappa-eq-CG}
\eea These also follow from the energy-conservation equation and the
trace part of ADM propagation equation in Eqs.\ (104), (102) of
\cite{NL}.

In \cite{second-order-CQG,second-order-PRD} we {\it identified} to
the second order \bea
   & & \delta \mu \equiv \delta \varrho, \quad
       \kappa \equiv - {1 \over a} \nabla \cdot {\bf u},
   \label{identify-second-order}
\eea where $\delta \varrho$ and ${\bf u}$ are Newtonian density and
velocity perturbations, respectively. As we ignore the rotational
mode, the velocity is of potential type with ${\bf u} = \nabla u$.
Apparently, we need $\chi$ to the linear order only, and to that
order we have \cite{second-order-CQG,second-order-PRD} \bea
   & & \nabla \chi = a {\bf u},
   \label{identify-chi}
\eea where we {\it assume} a flat Friedmann background world model.
With these identifications of the relativistic metric and
energy-momentum perturbation variables (these are equivalently
gauge-invariant combinations, see Sec.\ \ref{sec:two-gauges}) with
the Newtonian hydrodynamic variables, Eqs.\ (\ref{delta-eq-CG}),
(\ref{kappa-eq-CG}) give \bea
   & & \dot \delta + {1 \over a} \nabla \cdot {\bf u}
       = - {1 \over a} \nabla \cdot \left( \delta {\bf u} \right),
   \label{delta-eq-3rd} \\
   & & {1 \over a} \nabla \cdot \left( \dot {\bf u}
       + H {\bf u} \right)
       + 4 \pi G \mu \delta
       = - {1 \over a^2} \nabla \cdot \left( {\bf u}
       \cdot \nabla {\bf u} \right)
   \nonumber \\
   & & \quad
       - \dot C^{(t)\alpha\beta} \left( {2 \over a} \nabla_\beta
       u_\alpha
       + \dot C^{(t)}_{\alpha\beta} \right).
   \label{u-eq-3rd}
\eea By combining these we have \bea
   & & \ddot \delta + 2 {\dot a \over a} \dot \delta
       - 4 \pi G \mu \delta
       = - {1 \over a^2} {\partial \over \partial t}
       \left[ a \nabla \cdot \left( \delta {\bf u} \right) \right]
       + {1 \over a^2} \nabla \cdot \left( {\bf u}
       \cdot \nabla {\bf u} \right)
   \nonumber \\
   & & \quad
       + \dot C^{(t)\alpha\beta} \left( {2 \over a} \nabla_\beta u_\alpha
       + \dot C^{(t)}_{\alpha\beta} \right),
   \label{density-eq-3rd}
\eea which also follows from Eq.\ (\ref{NL-eq3}). Except for the
presence of the gravitational waves as source terms Eqs.\
(\ref{delta-eq-3rd})-(\ref{density-eq-3rd}) are valid {\it exactly}
in the Newtonian system \cite{Peebles-1980}.

Although our relativistic equations are valid to the second order,
the Newtonian equations are valid to fully nonlinear order. Thus,
all nonvanishing higher-order perturbation terms in the relativistic
case are pure general relativistic corrections. Recently, we have
presented such pure general relativistic correction terms appearing
in the third order perturbations in \cite{third-order}.

%%%%%%%%%%%%%%%%%%%%%%%%%%%%%%%%%%%%%%%%%%%%%%%%%%%%%%%%%%%%%%%
\subsection{The synchronous gauge}
                                                   \label{sec:SG}

The synchronous and comoving gauge conditions correspond to taking
\cite{LL} \bea
   & & v \equiv 0 , \quad
       \beta \equiv 0 ; \quad
       \alpha = 0.
   \label{SG}
\eea In this work, we call this simply the {\it synchronous} gauge.
Thus, we have $\dot \gamma = \chi/a^2$. If we take $v \equiv 0$ and
$\beta \equiv 0$ as the temporal and the spatial gauge conditions,
respectively, the momentum conservation equation gives $\alpha = 0$
to {\it all} orders in perturbations; although this is well known in
\cite{LL}, we give proofs in the Appendix B. Thus, in the
zero-pressure medium without rotation we can simultaneously impose
the comoving ($v = 0$) and the synchronous ($\alpha = 0$) temporal
gauge conditions as long as we also take $\beta \equiv 0$ as the
spatial gauge condition \cite{LL}; Kasai took these conditions in
his work in \cite{Kasai-1992}.

The original synchronous gauge used by Lifshitz \cite{Lifshitz-1946}
took $\alpha = 0$ and $\beta = 0$ as the temporal and the spatial
gauge conditions, respectively. These gauge conditions are known to
be {\it incomplete} in fixing both the temporal and the spatial
gauge modes even to the linear order. Thus, even after imposing
these gauge conditions we have remaining gauge modes present in the
solutions, in general. Meanwhile, $v \equiv 0$ and $\gamma \equiv 0$
fix the temporal and spatial gauge degree of freedoms completely,
thus no gauge mode is present in the solution, see Sec.\
\ref{sec:gauge-issue}. Since the original synchronous gauge implies
$v = 0$ (the nonvanishing solution of $v$ is the remnant temporal
gauge mode) in the zero-pressure case, we only have to pay attention
to the possible presence of the spatial gauge mode. In this gauge we
have $\tilde g_{00} = - a^2 = 1/\tilde g^{00}$ and $\tilde
g_{0\alpha} = 0 = \tilde g^{0\alpha}$. Thus, the fluid four-vector
in Eq.\ (\ref{u-pressureless}) becomes \bea
   & & \tilde u_0 = - a, \quad
       \tilde u_\alpha = 0; \quad
       \tilde u^0 = {1 \over a}, \quad
       \tilde u^\alpha = 0,
   \label{u-SG}
\eea which can be compared with Eq.\ (\ref{u-CG}) in the comoving
gauge. Thus, since $\tilde u^\alpha = 0$, our fluid four-vector
corresponds to the conventionally known comoving four-vector
\cite{Taub-1978}, and simultaneously normal because $\tilde u_\alpha
= 0$ as well. All the statements in the above two paragraphs are
valid for all perturbational orders.

Using Eq.\ (\ref{sigma}), Eqs.\ (\ref{NL-SG-eq1}), (\ref{NL-SG-eq2})
give \bea
   & & \dot \delta
       - \kappa
       = \delta \kappa,
   \label{delta-eq-SG} \\
   & & \dot \kappa
       + 2 H \kappa
       - 4 \pi G \mu \delta
   \nonumber \\
   & & \quad
       = \left( {1 \over a^2} \chi^{,\alpha\beta}
       + \dot C^{(t)\alpha\beta} \right)
       \left( {1 \over a^2} \chi_{,\alpha\beta}
       + \dot C^{(t)}_{\alpha\beta} \right).
   \label{kappa-eq-SG}
\eea These also follow from the energy-conservation equation and the
trace part of ADM propagation equation in Eqs.\ (104), (102) of
\cite{NL}. By combining these equations we have \bea
   & & \ddot \delta + 2 H \dot \delta - 4 \pi G \mu \delta
       = \dot \delta^2 + 4 \pi G \mu \delta^2
   \nonumber \\
   & & \quad
       + \left( {1 \over a^2} \chi^{,\alpha\beta}
       + \dot C^{(t)\alpha\beta} \right)
       \left( {1 \over a^2} \chi_{,\alpha\beta}
       + \dot C^{(t)}_{\alpha\beta} \right).
   \label{ddot-delta-eq-SG}
\eea Apparently, these equations in the synchronous gauge look
simpler than Eqs.\ (\ref{delta-eq-CG}), (\ref{kappa-eq-CG}) in the
comoving gauge. Compared with Eqs.\ (\ref{delta-eq-CG}),
(\ref{kappa-eq-CG}) in the comoving gauge, in Eqs.\
(\ref{delta-eq-SG}), (\ref{kappa-eq-SG}) we lack the
convective-derivative-like terms in the left-hand-sides. By changing
the time derivatives as \bea
   & & {\partial \over \partial t} \rightarrow
%       {\partial \over \partial t} - {1 \over a} N^\alpha
%       \nabla_\alpha =
       {\partial \over \partial t} + {1 \over a^2} ( \nabla \chi )
       \cdot \nabla
       = {\partial \over \partial t} + {1 \over a} {\bf u} \cdot
       \nabla,
   \label{time-derivative}
\eea we can show that Eqs.\
(\ref{delta-eq-SG})-(\ref{ddot-delta-eq-SG}) become the same ones in
the comoving gauge in Eqs.\ (\ref{delta-eq-CG}),
(\ref{kappa-eq-CG}), and (\ref{density-eq-3rd}); in the last step of
Eq.\ (\ref{time-derivative}) we used Eq.\ (\ref{identify-chi}) which
is valid for a flat background.

{\it If} we make the same identification of the density and velocity
perturbations as in Eqs.\ (\ref{identify-second-order}),
(\ref{identify-chi}), thus assuming a flat background, Eqs.\
(\ref{delta-eq-SG}), (\ref{kappa-eq-SG}) become: \bea
   & & \dot \delta
       + {1 \over a} \nabla \cdot {\bf u}
       = - {1 \over a} \delta \nabla \cdot {\bf u},
   \label{delta-eq-SG2} \\
   & & {1 \over a} \nabla \cdot
       \left( \dot {\bf u} + H {\bf u} \right)
       + 4 \pi G \mu \delta
       = - {1 \over a^2} \left( \nabla^\beta u^\alpha \right)
       \left( \nabla_\beta u_\alpha \right)
   \nonumber \\
   & & \quad
       - \dot C^{(t)\alpha\beta}
       \left( {2 \over a} \nabla_\beta u_\alpha
       + \dot C^{(t)}_{\alpha\beta} \right).
   \label{kappa-eq-SG2}
\eea Ignoring the gravitational waves, these equations can be
identified as the Newtonian hydrodynamic equations in the Lagrangian
frame.

Although equations in the synchronous gauge look simpler than the
ones in the comoving gauge, the presence of additional
convective-like terms in the comoving gauge allows us to make exact
(except for the gravitational waves) correspondence with the
Newtonian hydrodynamic equations in the Eulerian frame
\cite{second-order-CQG,second-order-PRD}. Whereas, the equations in
the synchronous gauge can be identified as the Newtonian equations
in the Lagrangian frame. However, the variables in the synchronous
gauge still have the remnant spatial gauge mode due to incomplete
fixing nature of the spatial gauge condition $\beta \equiv 0$ in
that gauge. That is, to the second order, $\delta$ and $\kappa$ in
the synchronous gauge have the remaining gauge modes, see Sec.\
\ref{sec:two-gauges}.

Now, we can relate the variables in the synchronous ($S$) gauge to
the ones in the comoving ($C$) gauge. {}From Eqs.\ (\ref{CG-SG}),
(\ref{chi-GT}), and (\ref{identify-chi}) we have \bea
   & & \delta_S = \delta_C + \left( \int^t {1 \over a^2} \nabla \chi dt
       + \nabla \gamma_{S,{\rm Gauge}} \right) \cdot \nabla
       \delta_C,
   \nonumber \\
   & &
       \kappa_S = \kappa_C + \left( \int^t {1 \over a^2} \nabla \chi dt
       + \nabla \gamma_{S,{\rm Gauge}} \right) \cdot \nabla
       \kappa_C,
   \label{SG-CG-sol}
\eea where $\gamma_{S,{\rm Gauge}}$ is the gauge mode present to the
linear order in $\gamma$; see the next section. In a flat
background, from Eq.\ (\ref{identify-chi}) we have $\nabla \chi = a
{\bf u}$. Notice that, even after ignoring the gauge modes
$\delta_S$ and $\kappa_S$ naturally differ from $\delta_C$ and
$\kappa_C$, respectively, because the final equations are different.
Using Eq.\ (\ref{SG-CG-sol}), Eqs.\
(\ref{delta-eq-SG})-(\ref{ddot-delta-eq-SG}) give Eqs.\
(\ref{delta-eq-CG}), (\ref{kappa-eq-CG}), and
(\ref{density-eq-3rd}).

Although the variables in the synchronous gauge have remnant spatial
gauge mode, somehow the equations in the synchronous gauge coincide
with the Newtonian ones in the Lagrangian frame. Meanwhile, the
Newtonian hydrodynamic equations have nothing to do with the gauge
mode which appears only in the relativistic treatment. We can show
that the situation is consistent in the synchronous gauge. {}From
Eqs.\ (\ref{xi_alpha-SG}), (\ref{GT-SG}) the gauge mode of
$\delta_{S,{\rm Gauge}} = \xi^\alpha \nabla_\alpha \delta_C$ is
proportional to the linear-order solution of $\delta_C$; similarly,
the gauge mode of $\kappa_{S,{\rm Gauge}} = \xi^\alpha \nabla_\alpha
\kappa_C$ is proportional to the linear-order solution of
$\kappa_C$. Thus, the behaviours of the gauge mode cannot be
distinguished from the solutions to the linear order, and can be
absorbed to the linear order solutions. We can also check that the
gauge modes in Eq.\ (\ref{SG-CG-sol}) cancel out in Eqs.\
(\ref{delta-eq-SG}) and (\ref{kappa-eq-SG}). In this sense, Eqs.\
(\ref{delta-eq-SG}) and (\ref{kappa-eq-SG}), thus Eqs.\
(\ref{ddot-delta-eq-SG}), (\ref{delta-eq-SG2}) and
(\ref{kappa-eq-SG2}) as well, are gauge-invariant. Therefore, to the
second order in the synchronous gauge, although the variables have
remnant gauge mode, the equations are gauge invariant; this happens
because the gauge mode temporally behaves exactly like one of the
physical solutions.

A similar situation occurs to the linear order in the original
synchronous gauge which took only $\alpha = 0 = \beta$
\cite{Lifshitz-1946}. Under these gauge conditions Lifshitz derived
\bea
   & & \ddot \delta + 2 H \dot \delta - 4 \pi G \mu \delta = 0,
   \label{ddot-delta-eq-SG-lin}
\eea which is the LHS of Eq.\ (\ref{ddot-delta-eq-SG}) and concides
with the later derived Newtonian equation \cite{Bonnor-1957}.
However, under these gauge conditions (i.e., without taking $v =
0$), $\delta$ still have the remnant gauge mode due to the
incomplete fixing nature of the temporal gauge condition $\alpha
\equiv 0$. It happens that the temporal behaviour of gauge mode of
$\delta$ is proportional to $H$ which {\it coincides} with one of
the two physical solutions, see \cite{GRG-1991}. Thus, although
$\delta$ has the gauge mode Eq.\ (\ref{ddot-delta-eq-SG-lin}) is
gauge invariant. In our synchronous gauge which also takes $v = 0$
the temporal gauge condition is fixed completely, but a similar
situation repeats due to an incomplete fixing nature of the spatial
synchronous gauge condition ($\beta \equiv 0$) now to the second
order in perturbation.

%%%%%%%%%%%%%%%%%%%%%%%%%%%%%%%%%%%%%%%%%%%%%%%%%%%%%%%%%%%%%%%
\section{Gauge issue}
                                                   \label{sec:gauge-issue}

%%%%%%%%%%%%%%%%%%%%%%%%%%%%%%%%%%%%%%%%%%%%%%%%%%%%%%%%%%%%%%%
\subsection{Gauge transformation}

Under a transformation between two coordinates $\hat x^a = x^a +
\tilde \xi^a$, the gauge transformation properties of all metric and
energy-momentum variables to the second order are presented in Sec.\
VI of \cite{NL}.

Since both the synchronous gauge and the comoving gauge take $v = 0$
we have \bea
   & & \tilde \xi^0 = 0.
\eea This follows from Eqs.\ (234) or (238) of \cite{NL}: in the
normal frame by setting $Q_\alpha = 0$ (i.e., $v = 0$) in both
gauges we have $\tilde \xi^0_{\;\;,\alpha} = 0$; or in the energy
frame, by setting $V_\alpha - B_\alpha + A B_\alpha + 2 V^\beta
C_{\alpha\beta} = 0$ (i.e., $\hat v = 0$) in both gauges, we again
have $\tilde \xi^0_{\;\;,\alpha} = 0$. Thus, without losing
generality we can take $\tilde \xi^0 = 0$.

To the second order, with $\tilde \xi^0 \equiv 0$, from Eqs.\ (229),
(232) of \cite{NL} we have \bea
   & & \hat \alpha
       = \alpha
       - \alpha_{,\alpha} \xi^\alpha
       - \beta_{,\alpha} \xi^{\alpha\prime}
       - {1 \over 2} \xi^{\alpha\prime} \xi_\alpha^\prime,
   \nonumber \\
   & &
       \hat \delta = \delta - \delta_{,\alpha} \xi^\alpha, \quad
       \hat \kappa = \kappa - \kappa_{,\alpha} \xi^\alpha,
   \label{GT1}
\eea where $\xi^\alpha \equiv \tilde \xi^\alpha$ with the index of
$\xi^\alpha$ based on $g^{(3)}_{\alpha\beta}$; a prime indicates the
time derivative based on the conformal time $\eta$ with $d \eta
\equiv d x^0 \equiv dt/a$. The gauge transformation property of
$\kappa$ follows from the scalar nature of the expansion scalar
$\tilde \theta$ with $\tilde \theta \equiv 3 H - \kappa$ where
$\tilde \theta$ is based on the normal frame; for the gauge
transformation of a scalar quantity, see Eq.\ (239) of \cite{NL}. To
the linear order, from Eq.\ (252) of \cite{NL} we have \bea
   & & \hat \beta_{,\alpha} = \beta_{,\alpha} + \xi_\alpha^\prime,
       \quad
       \hat \gamma_{,\alpha} = \gamma_{,\alpha} - \xi_\alpha.
   \label{GT2}
\eea Thus, $\chi \equiv a ( \beta + \gamma^\prime )$ is gauge
invariant to the linear order, and \bea
   & & \alpha - \alpha_{,\alpha} \gamma^{,\alpha}
       + {1 \over 2} \beta_{,\alpha} \beta^{,\alpha}, \quad
       \delta - \delta_{,\alpha} \gamma^{,\alpha}, \quad
       \kappa - \kappa_{,\alpha} \gamma^{,\alpha},
   \label{GI} \\
   & & \alpha - \alpha_{,\alpha} \gamma^{,\alpha}
       - \left( \beta + {1 \over 2} \gamma^\prime \right)^{,\alpha}
       \gamma^\prime_{,\alpha},
   \label{GI-alpha}
\eea are gauge invariant to the second order.

%%%%%%%%%%%%%%%%%%%%%%%%%%%%%%%%%%%%%%%%%%%%%%%%%%%%%%%%%%%%%%%
\subsection{Two gauges}
                                               \label{sec:two-gauges}

In the comoving gauge, by imposing $\gamma \equiv 0$ in all
coordinates (i.e., $\hat \gamma \equiv 0 \equiv \gamma$), from Eq.\
(\ref{GT2}) we have \bea
   & & \xi_\alpha = 0.
\eea Thus, the spatial gauge transformation property is fixed
completely. {}From Eq.\ (\ref{GT1}) we have \bea
   & & \hat \alpha = \alpha, \quad
       \hat \delta = \delta, \quad
       \hat \kappa = \kappa,
\eea and each variable in this gauge has unique gauge-invariant
counterpart as $\delta$ and $\kappa$ in Eq.\ (\ref{GI}) and $\alpha$
in Eq.\ (\ref{GI-alpha}). Thus, we can equivalently regard all
variables in this gauge as (spatially and temporally) gauge
invariant ones. {}For example, $\delta_{v,\gamma} \equiv \delta -
\delta_{,\alpha} \gamma^{,\alpha}$ is a unique gauge invariant
combination which is the same as $\delta$ in the $v = 0 = \gamma$
gauge conditions; for an explicit form of $\delta_{v,\gamma}$
including $v$, see Eq.\ (282) in \cite{NL}. We note that these
results (i.e., values remain the same in the comoving gauge
conditions, complete fixing of the gauge degrees of freedom, and
presence of unique corresponding gauge-invariant variables) continue
to be valid even in higher-order perturbations, \cite{NL}.

Whereas, in the synchronous gauge, by imposing $\beta \equiv 0$ in
all coordinates (i.e., $\hat \beta \equiv 0 \equiv \beta$), from
Eq.\ (\ref{GT2}) we have \bea
   & & \xi_\alpha^\prime = 0.
\eea Thus, even after imposing the gauge condition we have \bea
   & & \xi_\alpha = \xi_\alpha ({\bf x}),
   \label{xi_alpha-SG}
\eea which is the remaining gauge mode. Thus, under the synchronous
gauge, from Eqs.\ (\ref{GT1}), (\ref{GT2}) we still have \bea
   & & \hat \gamma_{,\alpha} = \gamma_{,\alpha} - \xi_\alpha,
   \nonumber \\
   & &
       \hat \alpha = \alpha - \alpha_{,\alpha} \xi^\alpha, \quad
       \hat \delta = \delta - \delta_{,\alpha} \xi^\alpha, \quad
       \hat \kappa = \kappa - \kappa_{,\alpha} \xi^\alpha,
   \label{GT-SG}
\eea where the transformation of $\gamma$ is valid to the linear
order. In this sense variables in the synchronous gauge have
remaining gauge modes even after imposing the gauge condition.
$\gamma$ has the remaining spatial gauge mode even in the linear
order, and the other variables have remaining gauge modes to the
second order.

%%%%%%%%%%%%%%%%%%%%%%%%%%%%%%%%%%%%%%%%%%%%%%%%%%%%%%%%%%%%%%%
\subsection{Transformation between the two gauges}

Using the gauge transformation properties of the variables we can
translate the equations and solutions in one gauge into the ones in
another gauge condition. We indicate the comoving gauge and the
synchronous gauge by subindices $C$ and $S$, respectively. To the
linear order we have \bea
   & & \beta_C = \gamma_S^\prime = {1 \over a} \chi.
   \label{chi-GT}
\eea We present three different ways to reach the transformation
properties.

{}First, we consider a transformation from the synchronous gauge
(unhat) to the comoving gauge (hat). {}From Eq.\ (\ref{GT1}) we have
\bea
   & & \alpha_C
       = - {1 \over 2} \xi^{\alpha\prime} \xi_\alpha^\prime,
   \nonumber \\
   & &
       \delta_C
       = \delta_S
       - \delta_{S,\alpha} \xi^\alpha, \quad
       \kappa_C
       = \kappa_S
       - \kappa_{S,\alpha} \xi^\alpha.
   \label{GT-SG-1}
\eea We need to determine the gauge transformation function
$\xi^\alpha$, apparently, only to the linear order. {}From Eq.\
(\ref{GT2}) we have \bea
   & & \xi_\alpha = \gamma_{S,\alpha}.
\eea Thus, Eq.\ (\ref{GT-SG-1}) becomes \bea
   & & \alpha_C
       = - {1 \over 2 a^2} \chi^{,\alpha} \chi_{,\alpha},
   \nonumber \\
   & &
       \delta_C
       = \delta_S
       - \delta_{S,\alpha} \gamma_S^{,\alpha}, \quad
      \kappa_C
       = \kappa_S
       - \kappa_{S,\alpha} \gamma_S^{,\alpha},
   \label{CG-SG}
\eea where we used Eq.\ (\ref{chi-GT}).

Second, we consider a transformation from the comoving gauge (unhat)
to the synchronous gauge (hat). {}From Eq.\ (\ref{GT1}) we have \bea
   & & \alpha_C
       = {1 \over 2} \xi^{\alpha\prime} \xi_\alpha^\prime
       + \beta_{C,\alpha} \xi^{\alpha\prime},
   \nonumber \\
   & &
       \delta_S
       = \delta_C
       - \delta_{C,\alpha} \xi^\alpha, \quad
       \kappa_S
       = \kappa_C
       - \kappa_{C,\alpha} \xi^\alpha.
   \label{GT-CG-1}
\eea {}From Eq.\ (\ref{GT2}) we have \bea
   & & \xi_\alpha = - \gamma_{S,\alpha}.
\eea Thus, Eq.\ (\ref{GT-CG-1}) leads to the same results in Eq.\
(\ref{CG-SG}).

{}Finally, the gauge invariant combination in Eq.\ (\ref{GI})
provides a simpler derivation. {}From the gauge invariance of
combinations in Eq.\ (\ref{GI}) we directly have Eq.\ (\ref{CG-SG}).

Using these gauge transformation properties in Eq.\ (\ref{CG-SG}) we
can derive Eqs.\ (\ref{delta-eq-SG}), (\ref{kappa-eq-SG}) from Eqs.\
(\ref{delta-eq-CG}), (\ref{kappa-eq-CG}), and vice versa.

%%%%%%%%%%%%%%%%%%%%%%%%%%%%%%%%%%%%%%%%%%%%%%%%%%%%%%%%%%%%%%%
%
% Discussion
%
%%%%%%%%%%%%%%%%%%%%%%%%%%%%%%%%%%%%%%%%%%%%%%%%%%%%%%%%%%%%%%%
\section{Discussion}

In this work we have compared the general relativistic weakly
nonlinear cosmological perturbation equations in two different gauge
conditions. In our previous works we have successfully shown that,
except for the coupling with gravitational waves, the relativistic
perturbation equations of a zero-pressure irrotational fluid
coincide exactly with the Newtonian ones to the second order in
perturbations. Such a relativistic-Newtonian correspondence was
available in our special comoving gauge condition in which all the
variables can be equivalently regarded as gauge invariant ones. In
this work we have compared these results with the ones in the
synchronous gauge. The case in the synchronous gauge was previously
studied without noticing the similarity or difference of the
equations with the Newtonian ones to the nonlinear orders.

In this work we compared equations in the synchronous gauge with the
ones in the comoving gauge and in the Newtonian case. Although the
variables in this gauge have remnant spatial gauge modes due to the
incomplete gauge fixing of the spatial gauge condition the equations
are gauge invariant. Ignoring the gravitational waves, the equations
in the synchronous gauge can be identified with the Newtonian
hydrodynamic equations in the Lagrangian frame to the second order,
whereas the equations in the comoving gauge can be identified as the
Newtonian ones in the Eulerian frame. These Eulerian and Lagrangian
correspondences can be understood because the fluid four-vector in
our comoving gauge is in fact normal as in Eq.\ (\ref{u-CG}) whereas
the four-vector in the synchronous gauge is both normal and comoving
(thus Lagrangian) as in Eq.\ (\ref{u-SG}). In our way to clarify the
case in the synchronous gauge we have addressed and resolved several
issues related to the two gauge conditions often to fully nonlinear
orders in perturbations.

%%%%%%%%%%%%%%%%%%%%%%%%%%%%%%%%%%%%%%%%%%%%%%%%%%%%%%%%%%%%%%%
%
% Acknowledgments
%
%%%%%%%%%%%%%%%%%%%%%%%%%%%%%%%%%%%%%%%%%%%%%%%%%%%%%%%%%%%%%%%
\subsection*{Acknowledgments}

We thank Professors J. Richard Bond, Lev Kofman and Misao Sasaki for
insightful and clarifying discussions. H.N. was supported by the
Korea Research Foundation Grant No. R04-2003-10004-0. J.H. was
supported by the Korea Research Foundation Grant No.
2003-015-C00253.

%%%%%%%%%%%%%%%%%%%%%%%%%%%%%%%%%%%%%%%%%%%%%%%%%%%%%%%%%%%%%%%
%
% Appendices
%
%%%%%%%%%%%%%%%%%%%%%%%%%%%%%%%%%%%%%%%%%%%%%%%%%%%%%%%%%%%%%%%
\section*{Appendices}

\subsection*{A. Invariance of fluid quantities}

Here, we {\it show} that the fluid quantities based on the fluid
four-vector in Eq.\ (\ref{u-pressureless}) do not depend on the
choice of $\tilde g^{0\alpha}$ (the spatial gauge condition) to all
orders in perturbations. Using a fluid four-vector $\tilde u_a$ the
energy-momentum tensor is decomposed into fluid quantities as
\cite{covariant,NL} \bea
   & & \tilde T_{ab} \equiv \tilde \mu \tilde u_a \tilde u_b
       + \tilde p \left( \tilde g_{ab} + \tilde u_a \tilde u_b \right)
       + \tilde q_a \tilde u_b + \tilde q_b \tilde u_a
       + \tilde \pi_{ab};
   \label{Tab} \\
   & & \tilde \mu \equiv \tilde T_{ab} \tilde u^a \tilde u^b, \quad
       \tilde p \equiv {1 \over 3} \tilde T_{ab} \tilde h^{ab}, \quad
       \tilde q_a \equiv - \tilde T_{cd} \tilde u^c \tilde h_a^d,
   \nonumber \\
   & &
       \tilde \pi_{ab} \equiv \tilde T_{cd} \tilde h_a^c \tilde h_b^d
       - \tilde p \tilde h_{ab},
   \label{fluid-Tab}
\eea where $\tilde h_{ab} \equiv \tilde g_{ab} + \tilde u_a \tilde
u_b$; we have $\tilde u^a \tilde q_a \equiv 0 \equiv \tilde u^a
\tilde \pi_{ab}$, $\tilde \pi_{ab} \equiv \tilde \pi_{ba}$, and
$\tilde \pi^a_a \equiv 0$. The variables $\tilde \mu$, $\tilde p$,
$\tilde q_a$ and $\tilde \pi_{ab}$ are the energy density, the
isotropic pressure (including the entropic one), the energy flux and
the anisotropic pressure (stress) based on the fluid four-vector,
respectively. Let us introduce another four-vector $\tilde U_a$ with
\bea
   & & \tilde U_0 = - a, \quad
       \tilde U_\alpha \equiv 0; \quad
       \tilde U^0 = {1 \over a}, \quad
       \tilde U^\alpha = - a \tilde g^{0\alpha}_U.
   \label{u-second}
\eea Thus, $\tilde U_a$ is subject to the same conditions as $\tilde
u_a$ in Eq.\ (\ref{u-pressureless}), but with possibly different
spatial gauge condition which could lead to $\tilde g^{0\alpha}_U
\neq \tilde g^{0\alpha}$. The fluid quantities based on $\tilde U_a$
are similarly defined as in Eqs.\ (\ref{Tab}), (\ref{fluid-Tab})
with $\tilde U_a$ replacing $\tilde u_a$: for example, we have
$\tilde \mu^U \equiv \tilde T_{ab} \tilde U^a \tilde U^b$, etc. We
can easily show that if $\tilde p = \tilde q_a = \tilde \pi_{ab} =
0$ we have \bea
   & & \tilde \mu^U
       \equiv \tilde T_{ab} \tilde U^a \tilde U^b
       = \tilde \mu \tilde u_a \tilde u_b \tilde U^a \tilde U^b
       = \tilde \mu \tilde u_0 \tilde u_0 \tilde U^0 \tilde U^0
       = \tilde \mu,
\eea and $\tilde p^U = \tilde q_a^U = \tilde \pi_{ab}^U = 0$, and
vice versa. This result is also valid to fully nonlinear order.

%%%%%%%%%%%%%%%%%%%%%%%%%%%%%%%%%%%%%%%%%%%%%%%%%%%%%%%%%%%%%%%
\subsection*{B. Justification of Eq.\ (\ref{SG})}

Here, we {\it show} that in a zero-pressure irrotational medium we
can take the original synchronous gauge ($\alpha \equiv 0 \equiv
\beta$) together with the temporal comoving gauge ($v \equiv 0$)
simultaneously to all orders in perturbations. This was known in
\cite{LL}, see Sec.\ 97 in \cite{LL}. Here, it is important to take
$\beta \equiv 0$ as the spatial synchronous gauge although we prefer
to take $\gamma \equiv 0$ as the spatial gauge condition because of
the remnant gauge mode in the $\beta \equiv 0$ case. We provide two
different proofs based on the ADM and the covariant formulations.

We begin by taking $v \equiv 0$ and $\beta \equiv 0$ as the temporal
and spatial gauge conditions, respectively. In Eq.\ (\ref{g^00}) we
showed that $\tilde g^{00} = - 1/N^2 = - 1/a^2$. The spatial gauge
condition $\beta = 0$ together with the irrotational condition
implies $\tilde g_{0\alpha} \equiv N_\alpha = 0$. Thus, from Eq.\
(2) of \cite{NL} we have $\tilde g_{00} \equiv - a^2 ( 1 + 2 \alpha
) = - N^2 = - a^2$. This implies that we have $\alpha = 0$.

Now, in the covariant approach, $v = 0$ and irrotational conditions
imply $\tilde u_\alpha = 0$. Since $\tilde u_a$ is the fluid
four-vector we take the energy frame, $\tilde q_a \equiv 0$. The
momentum conservation equation in Eq.\ (27) of \cite{NL} gives
$\tilde a_a = 0$. The spatial gauge condition $\beta = 0$ together
with the irrotational condition implies $\tilde g_{0\alpha} = 0$,
thus $\tilde g^{0\alpha} = 0$ as well. Since $\tilde a_\alpha \equiv
\tilde u_{\alpha;b} \tilde u^b = \tilde \Gamma^0_{0\alpha} = {1
\over 2} \tilde g^{00} \tilde g_{00,\alpha}$, $\tilde a_\alpha = 0$
implies that $\tilde g_{00}$ is a function of time only. Thus, we
have $\alpha = 0$.

If we impose $\alpha \equiv 0$ and $\beta \equiv 0$ as the gauge
condition, instead, we have non-vanishing $J_\alpha$ or $\tilde
u_\alpha$, thus nonvanishing $v$. In a zero-pressure medium this
nonvanishing $v$ can be identified as the remnant temporal gauge
mode, which can be set equal to zero without losing physical degree
of freedom.

%%%%%%%%%%%%%%%%%%%%%%%%%%%%%%%%%%%%%%%%%%%%%%%%%%%%%%%%%%%%%%%

%%%%%%%%%%%%%%%%%%%%%%%%%%%%%%%%%%%%%%%%%%%%%%%%%%%%%%%%%%%%%%%
\end{document}